\documentclass[conference]{IEEEtran}
\IEEEoverridecommandlockouts

\usepackage{graphicx}
\usepackage{subfigure}
\usepackage[hyphens]{url}
\usepackage{amsfonts}
\usepackage{booktabs}
\usepackage{romannum}
\usepackage{amsmath}
\usepackage{bm}
\usepackage{makecell}
\usepackage{color}
\usepackage{amsmath}
\usepackage{amssymb}
\usepackage{cite}
\usepackage{tcolorbox}

\def\BibTeX{{\rm B\kern-.05em{\sc i\kern-.025em b}\kern-.08em
    T\kern-.1667em\lower.7ex\hbox{E}\kern-.125emX}}
    
\begin{document}

\title{Deep Smart Contract Intent Detection}

\author{
    Youwei Huang$^{1,2}$, Sen Fang$^{3}$, Jianwen Li$^{2,4}$, Jiachun Tao$^{2,5}$, Bin Hu$^{6}$, and Tao Zhang$^{1\ast}$
    \\
    \normalsize $^{1}$ Macau University of Science and Technology, Macao SAR
    \\
    \normalsize $^{2}$ Institute of Intelligent Computing Technology, Suzhou, CAS, China
    \\
    \normalsize $^{3}$ North Carolina State University, USA
    \\
    \normalsize $^{4}$ Beijing Normal University - Hong Kong Baptist University United International College, China
    \\
    \normalsize $^{5}$ Suzhou City University, China
    \\
    \normalsize $^{6}$ Institute of Computing Technology, Chinese Academy of Sciences, China
    \\
    \normalsize devilyouwei@foxmail.com, tazhang@must.edu.mo
    \\
    \normalsize $^{\ast}$Corresponding author
}
\maketitle

\begin{abstract}
In recent years, research in software security has concentrated on identifying vulnerabilities in smart contracts to prevent significant losses of crypto assets on blockchains. Despite early successes in this area, detecting developers' intents in smart contracts has become a more pressing issue, as malicious intents have caused substantial financial losses. Unfortunately, existing research lacks effective methods for detecting development intents in smart contracts.

To address this gap, we propose \textsc{SmartIntentNN} (Smart Contract Intent Neural Network), a deep learning model designed to automatically detect development intents in smart contracts. \textsc{SmartIntentNN} leverages a pre-trained sentence encoder to generate contextual representations of smart contracts, employs a K-means clustering model to identify and highlight prominent intent features, and utilizes a bidirectional LSTM-based deep neural network for multi-label classification.

We trained and evaluated \textsc{SmartIntentNN} on a dataset containing over 40,000 real-world smart contracts, employing self-comparison baselines in our experimental setup. The results show that \textsc{SmartIntentNN} achieves an F1-score of 0.8633 in identifying intents across 10 distinct categories, outperforming all baselines and addressing the gap in smart contract detection by incorporating intent analysis.
\end{abstract}


\section{Introduction}
Web3, a term first coined by Gavin Wood\footnote{https://gavwood.com} within the Ethereum ecosystem~\cite{buterin2014next, wood2014ethereum, antonopoulos2018mastering}, refers to a decentralized network where applications, known as Decentralized Applications (DApps), run on blockchain infrastructure~\cite{nakamoto2008bitcoin, porru2017blockchain, bashir2017mastering, zheng2018blockchain}. 
Web3 enables DApps to operate in a trustless environment by leveraging blockchain’s immutable and distributed ledger technology.
A smart contract, which serves as the backbone of DApp development, is defined as a computer program and transaction protocol designed to automatically execute, control, or document legally relevant events and actions based on the terms of a contract or agreement~\cite{szabo1996smart, zou2019smart, ethereum2022smart}.

However, similar to other computer programs, smart contracts are susceptible to exploitation. They are exposed to vulnerabilities that hackers can exploit, as well as to malicious intents from developers aiming to defraud users. We categorize these risks into two types: external and internal. External risks arise from attacks outside the smart contract, typically involving hackers exploiting vulnerabilities. In contrast, internal risks stem from malicious code intentionally designed and embedded by developers inside the smart contracts.

\textbf{What are external risks?}
When considering smart contracts as computer programs, the primary external risks stem from the exploitation of vulnerabilities~\cite{he2020smart}. A notable example is the infamous DAO attack, triggered by a \textit{reentrancy vulnerability}~\cite{jiang2018contractfuzzer, su2021evil}, in which an attacker repeatedly invoked a function before the previous execution completed, enabling unauthorized withdrawals. Another common issue is \textit{unchecked exceptions}~\cite{luu2016making, brent2018vandal, kalra2018zeus}, wherein improper error handling leads to unexpected contract behavior. Additionally, \textit{integer overflow/underflow}~\cite{perez2019smart} can cause arithmetic operations to wrap around, leading to incorrect account balances or infinite loops. More vulnerabilities in smart contracts have been comprehensively summarized by \emph{Chu et al.}~\cite{chu2023survey}. These external risks are attacks from outside the smart contract that exploit its vulnerabilities.

\begin{figure}[ht]
    \centering\includegraphics[width=1\linewidth]{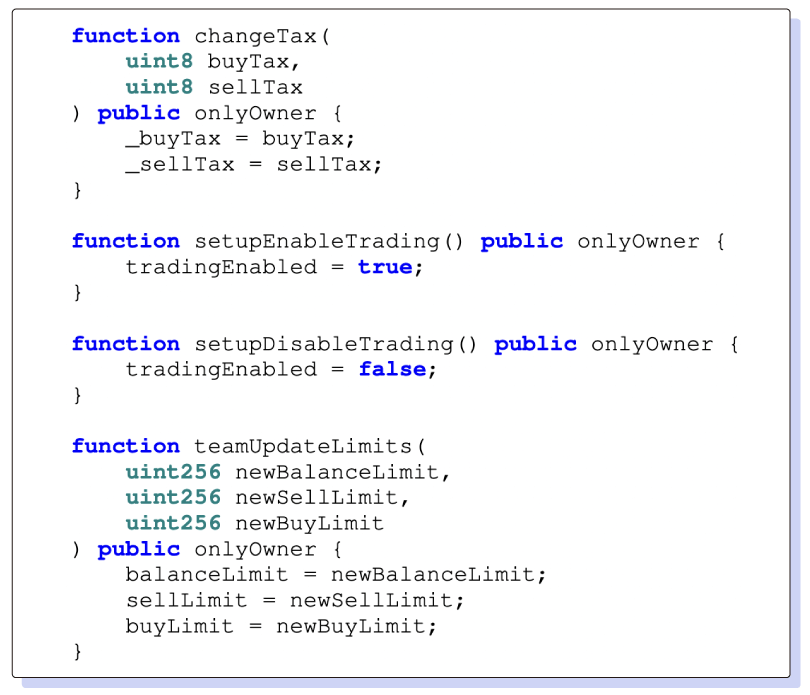}
    \caption{
        Examples of a smart contract with malicious intents.
        BSC address: 0xDDa7f9273a092655a1cF077FF0155d64000ccE2A.
    }
    \label{intentalriskem}
\end{figure}

\textbf{What are internal risks?}
In legal contracts, malicious terms can lead to significant losses for users. Similarly, in smart contracts, which are transaction protocols designed by developers, harmful terms can be embedded in the form of computer code. Our investigation reveals that, in recent years, an increasing number of risks have been intentionally injected by smart contract developers or DApp creators~\cite{torres2019art, hu2022scsguard, li2023siege, zhou2023dapphunter}. Malicious DApps purposely inject tricks and backdoors into their smart contracts to misappropriate users' funds~\cite{xia2021trade}.

As illustrated in Fig.~\ref{intentalriskem}, several examples of internal risks in a smart contract are shown. In these examples, all functions include the \textbf{\textit{onlyOwner}} modifier, which centralizes control in the hands of the contract owner. For instance, by adding the \textbf{\textit{onlyOwner}} modifier to the \textit{changeTax} function, the development team can arbitrarily change the tax fee for asset swaps. Similarly, \textit{teamUpdateLimits} with the \textbf{\textit{onlyOwner}} modifier allows developers to modify transaction limits. The other two functions are even more concerning, as they allow the owner to directly enable or disable trading within the smart contract. We define these risks, stemming from developers' intents, as internal or intentional risks.

\textbf{Our research} focuses on detecting internal risks in smart contracts by identifying negative developer intents embedded within the code. While there is extensive research on external risks, primarily in the area of smart contract vulnerability detection~\cite{chu2023survey}, few studies address internal risks. Moreover, no existing methods specifically target the detection of developer intents from the context of smart contract code. Currently, identifying such intents relies on expert manual audits, which are both time-consuming and costly~\cite{binance_smart_contract_audit}.

We present \textsc{SmartIntentNN}, a deep learning model comprising three key components: a Universal Sentence Encoder (USE)~\cite{cer2018universal} for contextual representation of smart contracts, a K-means clustering model~\cite{macqueen1967classification, krishna1999genetic} to highlight distinctive intent features, and a bidirectional long short-term memory (BiLSTM) network~\cite{hochreiter1997long, graves2005framewise2} for multi-label classification. Our model is implemented using TensorFlow.js~\cite{abadi2016tensorflow, smilkov2019tensorflow} and can detect ten categories of intents (see Table~\ref{tablelabels}).

We trained and evaluated \textsc{SmartIntentNN} on a dataset of over 40,000 smart contracts, comparing it against self-comparison baselines, including classic LSTM~\cite{hochreiter1997long}, BiLSTM~\cite{graves2005framewise2}, CNN~\cite{lecun1995convolutional} models, and generative large language models (LLMs) such as GPTs~\cite{achiam2023gpt}. \textsc{SmartIntentNN} achieved an \textit{F1-score} of 0.8633, an \textit{accuracy} of 0.9647, \textit{precision} of 0.8873, and \textit{recall} of 0.8406, surpassing all baselines.

\textbf{Our contributions} are summarized as follows:
\begin{itemize}
\item To the best of our knowledge, this study is the first to propose a method for detecting developer intents in smart contracts using deep learning techniques.
\item We have compiled a comprehensive dataset of over 40,000 labeled smart contracts, covering 10 categories of developer intents.
\item We have open-sourced \textsc{SmartIntentNN}, including all documentation, source code, datasets, and models. Please refer to \textcolor{blue}{\url{https://github.com/web3se-lab/web3-sekit}}.
\end{itemize}

\section{Motivation}
\textbf{Why is it essential to identify the developers' intent in smart contract development?} -- \textbf{For the love of money is the root of all evil.}
In traditional web software development, the concept of developers' intent is rarely discussed, as these applications are typically released by a trustworthy and centralized entity. These conventional applications make revenue by providing high-quality software services and do not rely on cryptocurrency systems. The entities behind these applications are responsible for the overall quality of the software.

In stark contrast, Web3 applications are decentralized and financially driven, allowing any developer to deploy a smart contract and promote it to users for financial purposes. 
More critically, smart contracts in Web3 applications often include their economic systems, which are tied to real-value cryptocurrencies (\textit{e.g.}, Bitcoin). 
This is where potential risks arise. As the assets held within a smart contract significantly increase in value, unscrupulous developers can exploit embedded malicious code to misappropriate users' cryptocurrencies.

Given the decentralized and value-driven nature of DApp development, it is crucial to detect and identify developers’ intents. Without such a software process, DApp users are at a heightened risk of exposure to malicious activities, which can result in substantial financial losses.

\textbf{Did developers' malicious intent cause financial damage?}
According to the 2022 Crypto Crime Report by Chainalysis~\cite{chainanlysis2022report}, cryptocurrency scammers have stolen approximately \$7.8 billion worth of cryptocurrency from victims, with over \$2.8 billion resulting from rug pulls. A rug pull occurs when developers invoke malicious functions embedded in smart contracts to perform unfair transactions, such as illegally withdrawing funds or removing liquidity pools to misappropriate crypto assets~\cite{binance2022rug, zhou2024stop}. Compared to data from 2020, the losses in 2021 surged by \$82\%. Further investigation by \textsc{HoneyBadger} revealed that 690 honeypot smart contracts accumulated more than \$90,000 in profit for their creators~\cite{torres2019art}.

The malicious intent of developers has indeed caused, and will likely continue to cause, significant financial damage to DApp users. Common schemes such as rug pulls, honeypots, and phishing scams exemplify this risk.

\textbf{Why do we propose an automated approach for detecting intent in smart contracts?}
Users need protection against smart contracts that conceal unsafe intents to prevent these intents from evolving into real harmful actions. Developers, on the other hand, often engage in practices like code cloning, reuse, and integrating third-party libraries during Web3 development, which can inadvertently introduce risky third-party code. An automated detection tool aids developers in identifying and avoiding such intents, ensuring the integrity and security of their projects.

From a cost perspective, traditional security audits for smart contracts are manually performed by experts, making them time-consuming and costly. Although existing research has introduced automated methods for analyzing smart contracts, these primarily focus on vulnerabilities or scams\cite{torres2019art, hu2022scsguard, chu2023survey}. An automated intent detection approach can significantly reduce the time, labor, and costs associated with security auditing. This motivates us to propose an automated method for detecting intents in smart contracts, thereby effectively reducing auditing costs and enhancing overall security.

\section{Background}
In this section, we provide essential background knowledge, covering smart contracts with malicious intents, sentence embedding techniques, and bidirectional LSTMs.

\subsection{Malicious Smart Contract Intent}
Smart contracts facilitate online interactions via blockchain networks, with Ethereum being the most widely used platform. It serves as an illustrative example of how developers with malicious intents can exploit these contracts to steal users' funds. Ethereum and similar blockchain platforms operate as transaction-based systems, where both deploying and invoking smart contracts require transactions. Users are represented by externally owned accounts (EOAs), necessary for paying gas fees in ether (ETH), Ethereum's native cryptocurrency.

\begin{figure}[ht]
\centering\includegraphics[width=\linewidth]{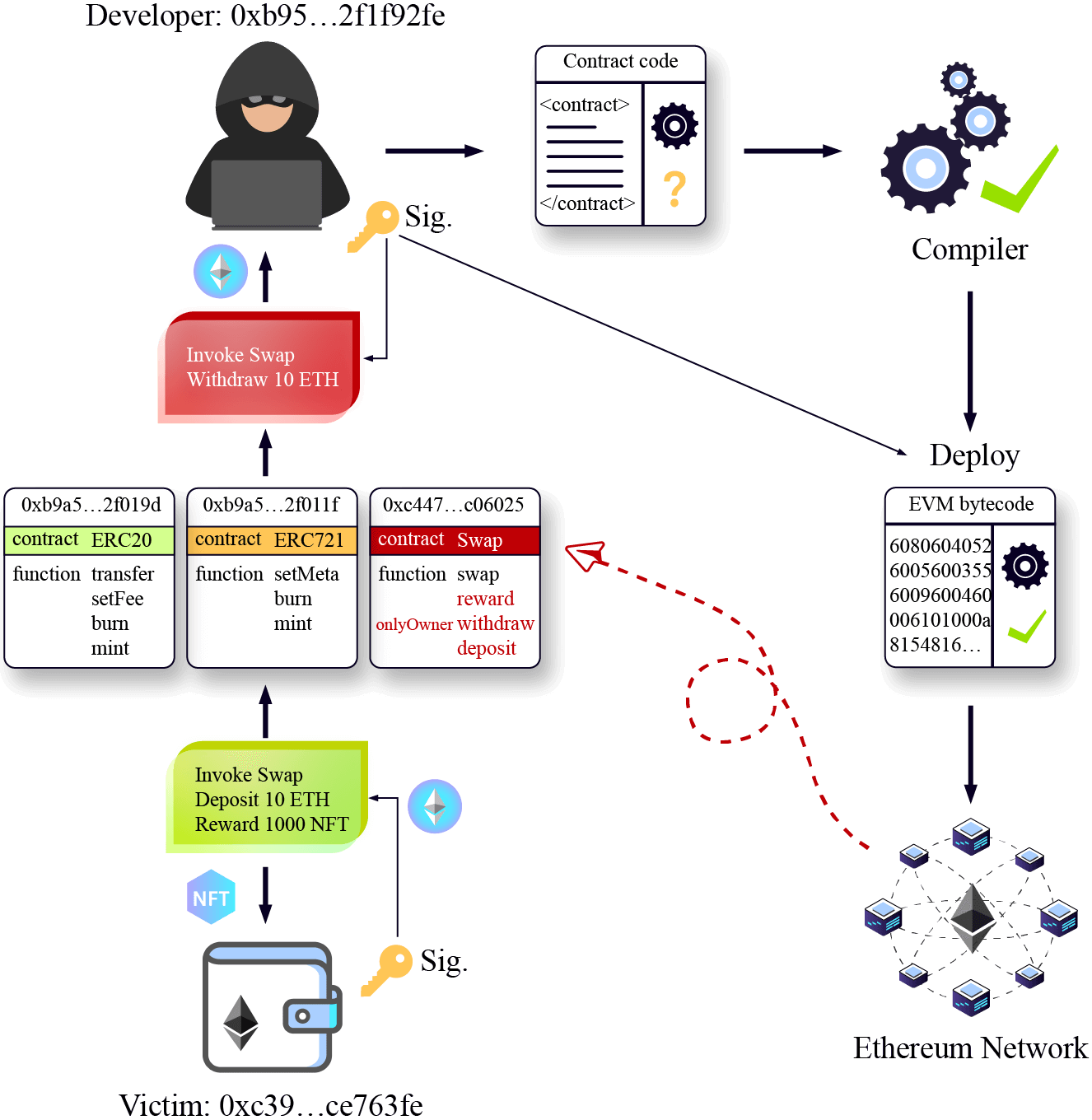}
\caption{
A depiction of how developers exploit smart contracts for illegal gain, illustrating the process of creating and deploying contracts with malicious intents.
}
\label{ethereum}
\end{figure}

As shown in Fig.~\ref{ethereum}, both developers and users maintain EOAs. A developer writes a smart contract, compiles it into bytecode executable on the Ethereum Virtual Machine (EVM)~\cite{ethereum_evm_2024}, and deploys it by signing a transaction with their private key, generating a unique contract address. Malicious developers might incorporate a \textit{reward} function into the smart contract to lure users, prompting them to stake ETH in exchange for incentives such as NFTs (non-fungible tokens)~\cite{wang2021non}.

A user might send a transaction to invoke a \textit{deposit} function, paying 10 ETH to the smart contract address. Subsequently, the user attempts to invoke the \textit{reward} function to receive NFTs. However, if the developer has secretly embedded a \textit{withdraw} function, they can immediately steal the user's 10 ETH. Once these transactions are confirmed on the blockchain, they cannot be altered or traced back, rendering the stolen funds irretrievable.

\subsection{Sentence Embedding}
Embedding, also known as distributed representation~\cite{mikolov2013distributed}, is a technique for learning dense representations of entities such as words, sentences, and images. Compared to word-level embeddings like word2vec~\cite{church2017word2vec} and GloVe~\cite{pennington2014glove}, the Universal Sentence Encoder (USE) provides sentence-level embeddings by aggregating word representations within a sentence~\cite{kiros2015skip}.

USE takes a sentence as input and outputs a contextual representation, ensuring that similar sentences are close in the generated vector space~\cite{mikolov2013distributed, mikolov2013efficient}. 
In this paper, we treat every \textit{function} snippet in a smart contract as a sentence and pass it into USE to obtain the contextual representation.

There are two main designs for USE. The first design is based on the transformer architecture~\cite{vaswani2017attention}, targeting high accuracy but at the cost of greater model complexity and resource consumption. The second design aims for efficient inference with slightly reduced accuracy by utilizing a deep averaging network (DAN)~\cite{iyyer2015deep}.

\subsection{Bidirectional LSTM}
A common LSTM cell is composed of a memory cell, a forget gate, an input gate, and an output gate~\cite{hochreiter1997long}. The memory cell retains values over arbitrary time intervals, while the gates regulate the flow of information into and out of the cell. A bidirectional LSTM (BiLSTM)~\cite{graves2005framewise2} incorporates two LSTM layers: one processes the input sequence in the forward direction and the other processes it in the backward direction. The output of a BiLSTM is the concatenation of the outputs from these two layers.

Compared to a standard LSTM, which only learns context dependency from the left side of the input sequence, a BiLSTM can learn dependencies from both sides of the input sequence. This bidirectional capability makes BiLSTM significantly more effective in understanding semantic contexts than a unidirectional LSTM.

\section{Dataset}
To train and evaluate \textsc{SmartIntentNN}, we compiled an extensive dataset of over 40,000 smart contracts from the Binance Smart Chain (BSC) explorer\footnote{\url{https://bscscan.com}}. BSC is an Ethereum-like blockchain, enabling smart contracts on BSC to also operate on Ethereum.

\begin{figure}[ht]
    \centering\includegraphics[width=\linewidth]{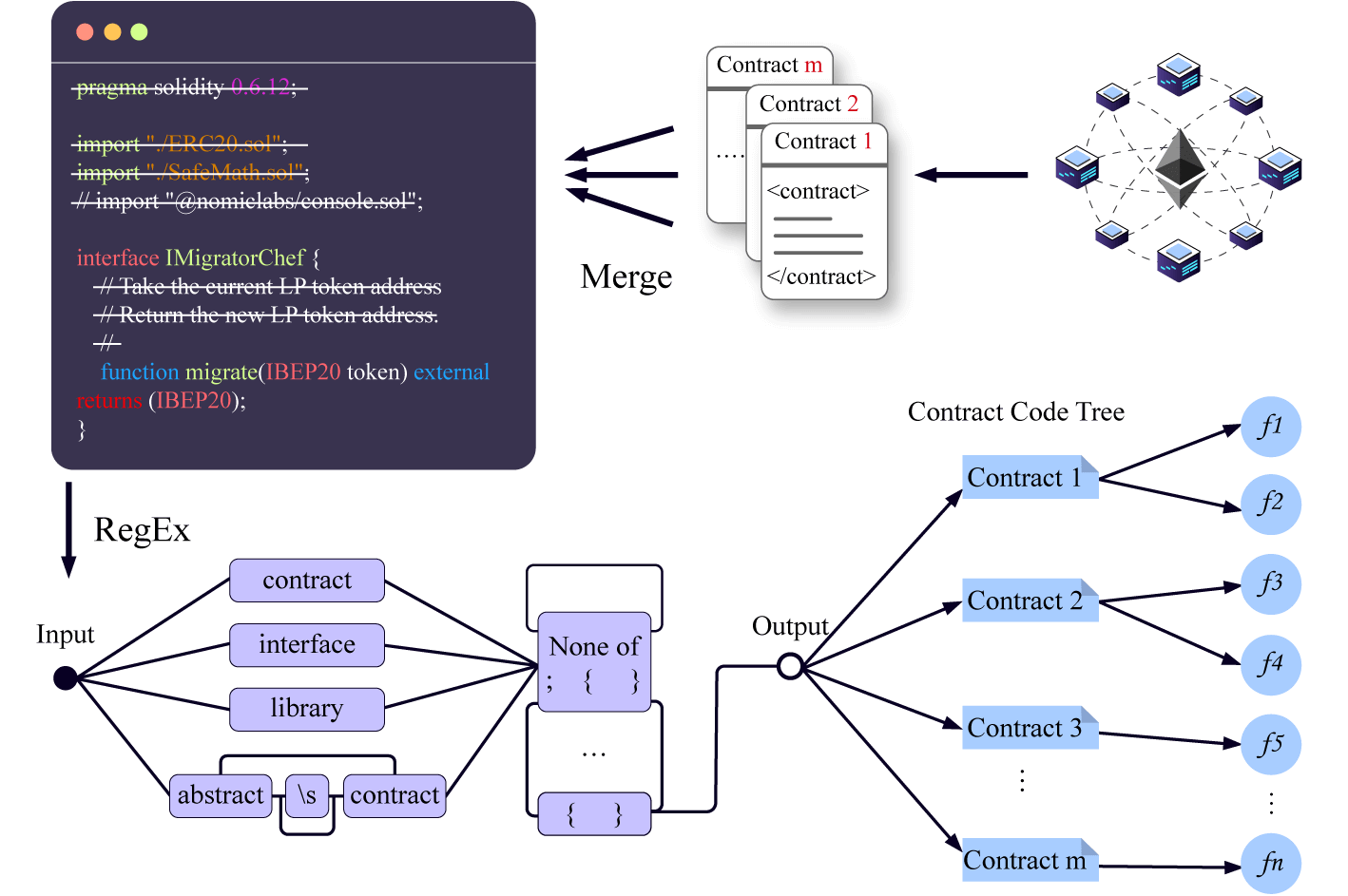}
    \caption{
        Dataset preprocessing steps: 
        (\rm{i}) download open-source smart contracts from the BSC blockchain and label them; 
        (\rm{ii}) merge and clean the source code; 
        (\rm{iii}) generate the smart contract code tree.
    }
    \label{dataset}
\end{figure}

We illustrate the data collection and preprocessing workflow in Fig.~\ref{dataset}. Initially, we downloaded a number of open-source smart contracts from the blockchain explorer. 
Then, we merged smart contracts that contained multiple files and cleaned redundant contracts.
Finally, we employed regular expressions (RegEx) to extract code snippets based on Solidity syntax keywords, generating a structured smart contract code tree.

\subsection{Intent Labels}
The dataset includes ten categories of unsafe intents in smart contracts, as detailed in Table~\ref{tablelabels}. These ten common intents were collected from the experiences of DApp developers but do not represent all potential categories of intents. The smart contracts in our dataset were labeled by experts, developers, auditors, and through various blockchain security websites, e.g., StaySafu\footnote{\url{https://www.staysafu.org}}.

\begin{table}[ht]
    \centering
    \caption{Categories of Intents}
    \begin{tabular}{crrr}
        \toprule
        \textbf{Id} & \textbf{Intent} & \textbf{Percentage} & \textbf{Instance}                        \\
        \midrule
        1           & Fee             & $26.86\%$           & \textit{setFeeAddress(address)}          \\
        2           & DisableTrading  & $5.34\%$            & \textit{enableTrading(bool)}             \\
        3           & Blacklist       & $3.82\%$            & \textit{require(!isBlacklisted[sender])} \\
        4           & Reflect         & $37.50\%$           & \textit{tokenFromReflection(uint256)}    \\
        5           & MaxTX           & $13.76\%$           & \textit{setMaxTxPercent(uint256)}        \\
        6           & Mint            & $8.53\%$            & \textit{mint(uint256)}                   \\
        7           & Honeypot        & $0.23\%$            & \textit{require(allow[from])}            \\
        8           & Reward          & $3.37\%$            & \textit{updateDividendTracker(address)}  \\
        9           & Rebase          & $0.53\%$            & \textit{LogRebase(uint256, uint256)}     \\
        10          & MaxSell         & $0.05\%$            & \textit{setMaxSellToken(uint256)}        \\
        \bottomrule
    \end{tabular}
    \label{tablelabels}
\end{table}

The ten categories of intents in Table~\ref{tablelabels} are described below:
\begin{itemize}
    \item[1] \textbf{Fee}: Arbitrarily changes transaction fees, directing them to specified wallet addresses.
    \item[2] \textbf{DisableTrading}: Enables or disables trading actions within a smart contract.
    \item[3] \textbf{Blacklist}: Restricts specified users' activities, potentially infringing on their trading rights.
    \item[4] \textbf{Reflect}: Redistributes transaction fees to holders based on their holdings, often used to incentivize holding native tokens.
    \item[5] \textbf{MaxTX}: Limits the maximum number or volume of transactions allowed.
    \item[6] \textbf{Mint}: Issues new tokens, potentially in an unlimited or controlled manner.
    \item[7] \textbf{Honeypot}: Traps user-provided funds by falsely promising to release funds while keeping the user's funds inaccessible.
    \item[8] \textbf{Reward}: Provides users with crypto assets as rewards to encourage token use, often regardless of the actual value of the rewards.
    \item[9] \textbf{Rebase}: Adjusts the total supply of tokens algorithmically to stabilize or change the token's price.
    \item[10] \textbf{MaxSell}: Limits the amount or frequency of token sales for specified users to restrict liquidity.
\end{itemize}

Each of these intents can appear multiple times in one smart contract.
Our statistical analysis revealed that the most frequent intent is \textbf{Reflect}, appearing in $33,268$ instances out of our dataset, accounting for $37.5\%$ of all intents. 
This is followed by \textbf{Fee} at $26.86\%$ and \textbf{MaxTX} at $13.76\%$. The intent with the least occurrence is \textbf{MaxSell}, appearing in only $68$ instances and accounting for a mere $0.05\%$. These statistics indicate a higher prevalence of certain risks, such as \textbf{Reflect}, \textbf{Fee}, and \textbf{MaxTX}, in smart contracts on the BSC.

It is important to note that the unsafe intents in our dataset are not equal to malicious intents. Instead, they represent potential internal risks in smart contracts. Only when developers exploit these intents to initiate unfair transactions do they become true malicious intents.

\subsection{Code Cleaning}
The source code of smart contracts on BSC exists in two forms: single-file and multi-file. A single-file contract has its \textit{import} contracts merged by developers using a flattener tool before uploading. For multi-file contracts, we merge all files into a single document. We remove the Solidity compiler version specification (\textit{pragma}), \textit{import} statements, and \textit{comments}, as \textit{pragma} does not convey any developer's intent, \textit{comments} do not affect intent implementation, and \textit{import} statements are redundant after merging contracts.

\subsection{Smart Contract Code Tree}
Smart contracts, being composed of code, cannot be directly fed into a neural network. To prepare the input data, we create a Smart Contract Code Tree (CCTree). The CCTree organizes the smart contract source code into three layers: (i) the root layer, representing the entire smart contract document; (ii) the contract layer, representing the \textit{contract} classes within the smart contract; (iii) the function layer, representing the \textit{function} contexts within each \textit{contract} class. The root layer contains a single node, denoted by $\mathbf{T}$. The second layer lists the \textit{contract} classes (e.g., \textit{contract}, \textit{interface}, \textit{library}, \textit{abstract contract}), denoted by $\mathbf{C}$. The third layer includes leaf nodes with code snippets associated with keywords such as \textit{function}, \textit{event}, \textit{modifier}, and \textit{constructor}, denoted by $\mathcal{F}$. This structure allows systematic access to code data for model training and evaluation.

\section{Approach}

\begin{figure*}[ht]
    \centering\includegraphics[width=\linewidth]{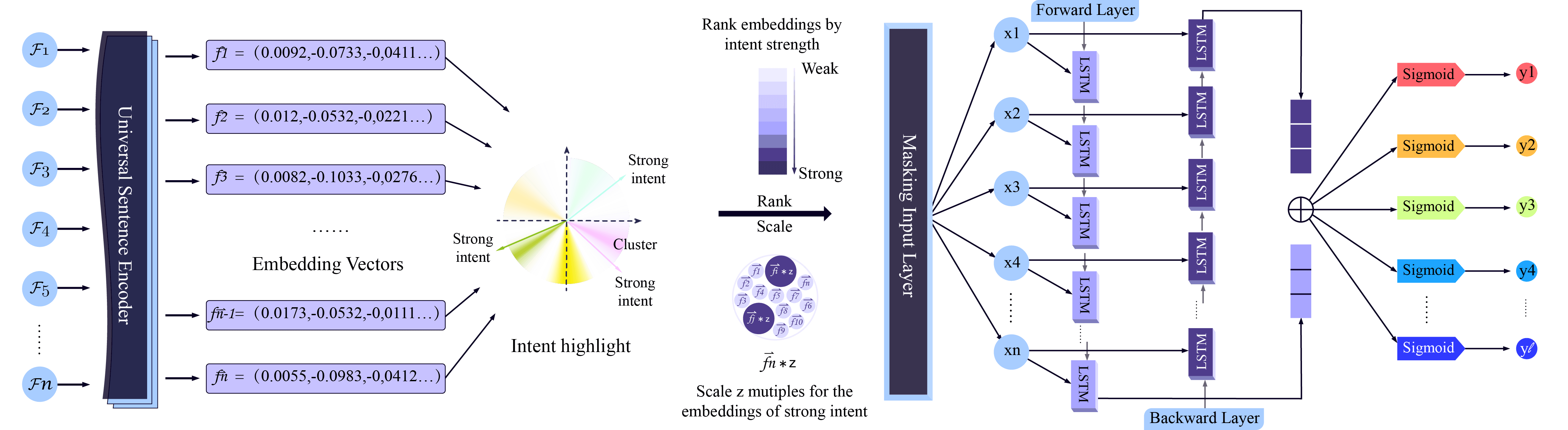}
    \caption{
    Overview of the \textsc{SmartIntentNN} workflow:
    (\rm{i}) encode smart contracts through the Universal Sentence Encoder;
    (\rm{ii}) identify and highlight features of developers' distinct intents in smart contracts using a K-means model;
    (\rm{iii}) feed the intent-highlighted data into a DNN for learning the representations of smart contracts.
    The architecture of our DNN includes an input layer, a BiLSTM layer, and a dense layer to output the multi-label binary classification results.
    }
    \label{framework}
\end{figure*}

In this section, we introduce the \textsc{SmartIntentNN} model and its approach for detecting intents in smart contracts. As illustrated in Fig.~\ref{framework}, \textsc{SmartIntentNN} comprises three core components. First, the context of smart contracts is encoded into embedding vectors using USE. Next, an intent highlight model with K-means clustering emphasizes features with strong intents. Finally, the highlighted data is fed into a BiLSTM-based DNN layer to capture smart contract representations and perform multi-label classification.

\subsection{Smart Contract Embedding}
To generate contextual representations of a smart contract, we begin by considering each \textit{function} in it.
Each \textit{function} is individually encoded, and their embeddings are then combined to create a holistic representation of the smart contract.

\begin{equation}
    \Phi(\mathcal{F}): \mathcal{F} \rightarrow \bm{f}
    \label{formula_embedding}
\end{equation}

Using the CCTree introduced in Section IV, an exhaustive traversal of all leaf nodes is performed. Consequently, each \textit{function} (denoted as $\mathcal{F}$, referring to a leaf node) in a smart contract is transformed into a sentence embedding using the pre-trained USE based on the DAN. This embedding process is encapsulated by Formula \ref{formula_embedding}, where $\Phi$ represents the contextual encoder, $\mathcal{F}$ is the \textit{function} context, and $\bm{f}$ is the resulting \textit{function} embedding vector.

To generate a sentence embedding for $\mathcal{F}$, the DAN model proceeds as follows:

\begin{equation}
    \mathrm{T}(\mathcal{F}): \mathcal{F} \rightarrow \mathbf{W}^\mathcal{F} = \{w_1, w_2, \dots, w_n\}
    \label{formula_token}
\end{equation}
\begin{equation}
    \phi(w_i): w_i \rightarrow \bm{w_i}, \quad w_i \in \mathbf{W}^\mathcal{F}
    \label{formula_word2vec}
\end{equation}
\begin{equation}
    \bm{f^0} = \frac{1}{n} \sum_{i=1}^n \bm{w_i}
    \label{formula_average}
\end{equation}

The series of operations from Formula \ref{formula_token} to Formula \ref{formula_average} represents the initial step in the DAN model, termed ``tokenizing". Here, $\mathrm{T}$ in Formula \ref{formula_token} transforms a sentence into word-level tokens. The text within \textit{function} $\mathcal{F}$ is dissected into an ordered set of word tokens $\mathbf{W}^\mathcal{F} = \{w_i\}_{i=1}^n$. Subsequently, $\phi$ in Formula \ref{formula_word2vec} converts these tokens into word embeddings $\{\bm{w_1}, \bm{w_2}, \dots, \bm{w_n}\}$. Following this, Formula \ref{formula_average} computes the average of all word embedding vectors to yield $\bm{f^0}$.

Deep feed-forward neural networks help learn increasingly abstract representations of input data with each layer~\cite{bengio2013representation}. Therefore, in the second step of the DAN model, the mean output $\bm{f^0}$ undergoes further enhancement through multiple feed-forward layers. Assuming $n$ feed-forward layers, each layer can be represented by Formula \ref{formula_forward}, where $\bm{W_i}$ is a weight matrix $\in \mathbb{R}^{k \times k}$ (with $k$ representing the dimension of vector $\bm{f^i}$), $\sigma$ is the activation function (e.g., $\mathrm{sigmoid}$ or $\mathrm{tanh}$), and $\bm{b_i}$ is the bias term.

\begin{equation}
    \bm{f^i} = \sigma(\bm{W_i} \cdot \bm{f^{i-1}} + \bm{b_i}), \quad i \in \{1, 2, \cdots, n\}
    \label{formula_forward}
\end{equation}

In the final step, $\bm{f^n}$ is fed into a $\mathrm{softmax}$ layer, generating a universal representation of the \textit{function} $\mathcal{F}$. Here, $\bm{W_s}$ is a weight matrix $\in \mathbb{R}^{m \times k}$, where the input size is $k$, and the output is a \textit{function} embedding vector with $m$ features. Formula \ref{formula_senclass} produces $\bm{f}$, which embeds the context of the \textit{function}.

\begin{equation}
    \bm{f} = \mathrm{softmax}(\bm{W_s} \cdot \bm{f^n} + \bm{b_s})
    \label{formula_senclass}
\end{equation}

These operations are applied to each \textit{function} in a smart contract. The resulting embedding vectors $\bm{f}$ are assembled into a matrix $\bm{X}$, which represents the entire smart contract, where $\bm{X} \in \mathbb{R}^{n \times m}$. Here, $n$ corresponds to the number of \textit{functions} in the smart contract, while $m$ represents the dimension of $\bm{f}$.

\subsection{Intent Highlight}
In a smart contract, not all \textit{functions} express the developer's intent. To enhance the model's understanding of these intent features, we introduce an intent highlight model to emphasize intent-related \textit{functions}. This model utilizes K-means clustering to predict the intent strength of each vector $\bm{f}$ in $\bm{X}$. By quantifying intent strength, we amplify the features of vectors that strongly reflect the developer's intent. This mechanism is represented by the formula \ref{formula_highlight}, where $\mathrm{H}$ denotes the K-means-based intent highlight model, and $\bm{X'}$ is the output of the intent-highlighted data.

\begin{equation}
    \bm{X'} = \mathrm{H}(\bm{X})
    \label{formula_highlight}
\end{equation}

To find the appropriate $k$ value for K-means clustering, we compute the occurrence rate of each \textit{function} $\mathcal{F}$ within a randomly selected subset of the entire smart contract dataset. This subset, $\mathbf{S}$, consists of $m$ smart contracts, denoted as $\mathbf{S} = \left\{\mathbf{T_i}\right\}_{i=1}^m$. The occurrence rate, $\mathrm{R}\left(\mathcal{F}\right)$, is defined as the frequency with which a \textit{function} appears in the subset, as detailed in Formula~\ref{formula_rate}. If a \textit{function} $\mathcal{F}$ exists within the smart contract tree $\mathbf{T_i}$, it is counted as $\mathbb{I}\left\{ \mathcal{F} \in \mathbf{T_i} \right\}$. We specifically tally the number of $\mathcal{F}$ \textit{functions} whose occurrence rate $\mathrm{R}\left(\mathcal{F}\right) > \rho$, where $\rho$ is an experimentally determined threshold. The total count of these high-frequency \textit{functions} ($\mathrm{R}\left(\mathcal{F}\right) > \rho$) gives us the value of $k$.

\begin{equation}
    \mathrm{R}\left(\mathcal{F}\right) = \frac{1}{m} \sum_{i=1}^m \mathbb{I}\left\{ \mathcal{F} \in \mathbf{T_i} \right\}
    \label{formula_rate}
\end{equation}

We compute the cosine distance between their embedding vectors when comparing the similarity of two documents~\cite{rahutomo2012semantic}. Consider two \textit{functions} ($\mathcal{F}_A$ and $\mathcal{F}_B$) with embedding vectors $\bm{f^A}=[f_1^A, f_2^A, \dots, f_n^A]$ and $\bm{f^B}=[f_1^B, f_2^B, \dots, f_n^B]$. Formula~\ref{formula_similarity} calculates the text similarity between $\mathcal{F}_A$ and $\mathcal{F}_B$, which is the cosine value of $\bm{f^A}$ and $\bm{f^B}$. We use Formula~\ref{formula_distance} to transform cosine similarity into cosine distance.

\begin{equation}
    \cos\left\langle\bm{f^A},\bm{f^B}\right\rangle = \frac{\bm{f^A} \cdot \bm{f^B}}{\left\|\bm{f^A}\right\| \left\|\bm{f^B}\right\|}
    \label{formula_similarity}
\end{equation}
\begin{equation}
    \mathrm{D}\left(\bm{f^A},\bm{f^B}\right) = 1 - \cos\left\langle\bm{f^A},\bm{f^B}\right\rangle
    \label{formula_distance}
\end{equation}

To identify the centroids for K-means clustering, we start with a training dataset $\left\{\bm{f'_i}\right\}_{i=1}^n$, where each $\bm{f'_i}$ is an embedding vector representing a \textit{function} $\mathcal{F}$ from the previously described subset $\mathbf{S}$. We initialize $k$ random centroids $\left\{\bm{c_j^t}\right\}_{j=1}^k$ for $t=0,1,\dots,z$, where $z$ is the maximum number of iterations. K-means iteratively updates centroids by minimizing the cosine distance, as opposed to the original Euclidean distance. Each \textit{function} embedding $\bm{f'_i}$ is assigned to the closest centroid $\bm{c_j^t}$, forming a set $\mathbf{M_j^t}$ for each centroid. The centroid $\bm{c_j^{t+1}}$ is updated based on the weighted average of all vectors in $\mathbf{M_j^t}$, as shown in Formula~\ref{formula_centroid}.

\begin{equation}
\bm{c_j^{t+1}} = \frac{\sum_{\bm{f'_i} \in \mathbf{M_j^t}} \bm{f'_i}}{\left|\mathbf{M_j^t}\right|}
\label{formula_centroid}
\end{equation}

The objective is to minimize the total within-cluster variation (TWCV), as depicted in Formula~\ref{formula_min_twcv}. If $\bm{f'_i}$ belongs to $\mathbf{M_j^t}$, the cosine distance is calculated. Here, $\mathrm{D}(\bm{f'_i}, \bm{c_j^t})$ represents the cosine distance between the \textit{function} vector $\bm{f'_i}$ and the centroid $\bm{c_j^t}$ at iteration $t$.

\begin{equation}
\min_{t=0}^{z} \sum_{j=1}^k \sum_{\bm{f'_i} \in \mathbf{M_j^t}} \mathrm{D}(\bm{f'_i}, \bm{c_j^t})
\label{formula_min_twcv}
\end{equation}

After each iteration, the centroid $\bm{c_j^t}$ is updated to $\bm{c_j^{t+1}}$. The training process continues until $t$ reaches the maximum number of iterations, $z$, or the reduction in TWCV becomes negligible. Ultimately, this process identifies the most appropriate centroids $\left\{\bm{c_j}\right\}_{j=1}^k$ for K-means, representing the spatial cluster centers where frequently occurring contract \textit{functions} are densely grouped.

\begin{figure}[ht]
    \centering\includegraphics[width=\linewidth]{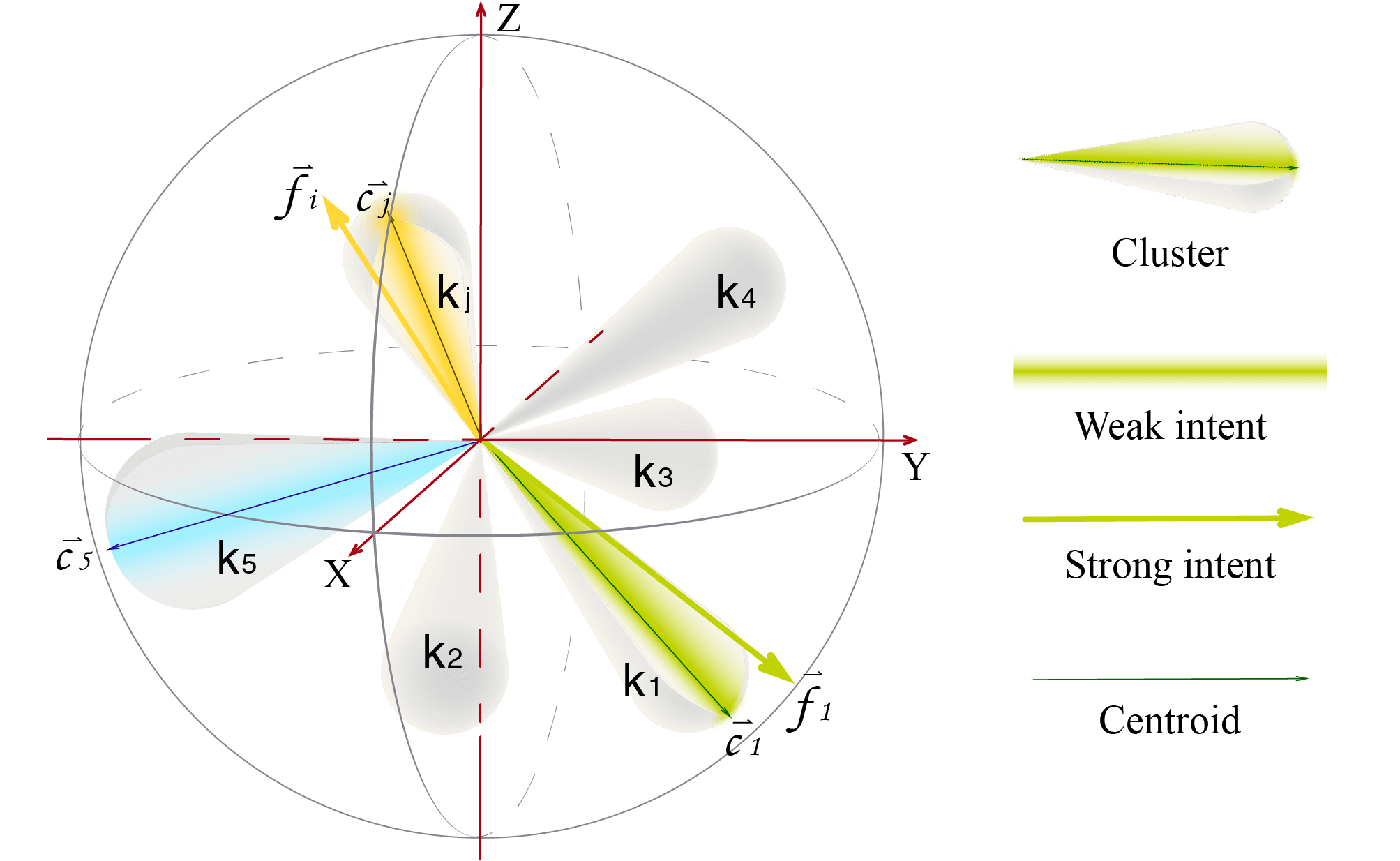}
    \caption{
        A 3D coordinate system illustrating the principle of intent highlighting. The larger the vector angle deviates from the centroid, the stronger its intent.
    }
    \label{intenthighlight}
\end{figure}

Once the K-means model has been trained, we can input any \textit{function} embedding vector $\bm{f_i}$ to calculate its within-cluster distance. As illustrated in Fig.~\ref{intenthighlight}, \textit{functions} that are closer to their centroids typically exhibit weaker developer intent. These are often frequently occurring and similar \textit{function} code snippets, commonly originating from public libraries, mainstream algorithms, or widely reused \textit{functions}. In contrast, \textit{functions} that are further from the centroid, known as centroid-outlying \textit{functions}, tend to be distinctive and express stronger intent. Here, we utilize the within-cluster distance as a measure to quantify the strength of a \textit{function}'s intent.

\begin{equation}
    \mathrm{H_{\mu}}\left(\bm{X}\right)\ = \bm{X} \odot \left(1 + (\mu - 1) \cdot \mathbb{I}\left\{ \mathrm{D}\left(\bm{f_i}, \bm{c_j}\right) \geq \lambda \right\}\right)
\label{formula_scale}
\end{equation}

Next, we proceed to effectively highlight the intent features within smart contracts. Using the K-means model, we predict the within-cluster distance for each $\bm{f_i}$ in $\bm{X}$ and scale the features of selected \textit{functions}, resulting in a new matrix $\bm{X'} \in \mathbb{R}^{n \times m}$. In Formula~\ref{formula_scale}, $i \in \{1, 2, \cdots, n\}$ and $j \in \{1, 2, \cdots, k\}$, where $\lambda$ denotes a distance threshold and $\mu$ represents the scaling factor applied to $\bm{f_i}$ if its within-cluster distance $\mathrm{D}(\bm{f_i},\bm{c_j}) \geq \lambda$. These transformations are visually represented in the central portion of Fig.~\ref{framework}, adjacent to the intent highlight model.

\subsection{Multi-label Classification}
In this section, we introduce the final part of Fig.~\ref{framework}, which involves using a DNN for multi-label classification. The architecture of the DNN comprises three layers: an input layer, a BiLSTM layer, and an output layer.
We input the intent-highlighted data, represented by the matrix $\bm{X'}$ produced by the previous intent highlight model, into the DNN.

Initially, the data is fed into the input layer, which accepts a sequence of $\mathbb{R}^{p \times m}$. Here, $p$ corresponds to the number of \textit{functions} $\mathcal{F}$ input at each timestep, and $m$ represents the dimensionality of each \textit{function} embedding $\bm{f_i}$. For instance, USE outputs a vector of 512 dimensions, then $m=512$. The row size of $\bm{X'}$ may vary depending on the number of \textit{functions} present in each smart contract. In cases where the number of rows in $\bm{X'}$ is less than $p$, additional rows are padded with zero vectors. This input layer also functions as a masking layer, allowing TensorFlow to skip over these padded timesteps by setting the masking value to zero~\cite{tensorflow2024latest}.

Next, the data flows into the BiLSTM layer, which also accepts a $\mathbb{R}^{p \times m}$ matrix, denoted by $\bm{X''} = [\bm{f_i}]_{i=1}^p$, output from the input layer. Each LSTM layer in the BiLSTM contains $p$ memory cells, thus there are a total of $2p$ cells when considering both forward and backward layers. Each row vector $\bm{f_i}$ from $\bm{X''}$ is input into corresponding cells in both the forward and backward layers.

\begin{equation}
    \mathrm{G}^\mathsf{f}_i=\alpha\left(\bm{W^\mathsf{f}_i}\bm{f_i}+\bm{U^\mathsf{f}_i}\bm{h_{i-1}}+\bm{b^\mathsf{f}_i}\right)
    \label{formula_lstm_1}
\end{equation}
\begin{equation}
    \mathrm{G}^\mathsf{u}_i=\alpha\left(\bm{W^\mathsf{u}_i}\bm{f_i}+\bm{U^\mathsf{u}_i}\bm{h_{i-1}}+\bm{b^\mathsf{u}_i}\right)
    \label{formula_lstm_2}
\end{equation}
\begin{equation}
    \mathrm{G}^\mathsf{o}_i=\alpha\left(\bm{W^\mathsf{o}_i}\bm{f_i}+\bm{U^\mathsf{o}_i}\bm{h_{i-1}}+\bm{b^\mathsf{o}_i}\right)
    \label{formula_lstm_3}
\end{equation}
\begin{equation}
    \widetilde{\Theta}_i=\beta\left(\bm{W^\theta_i}\bm{f_i}+\bm{U^\theta_i}\bm{h_{i-1}}+\bm{b^\theta_i}\right)
    \label{formula_lstm_4}
\end{equation}
\begin{equation}
    \Theta_i=\mathrm{G}^\mathsf{f}_i\odot\Theta_{i-1}+\mathrm{G}^\mathsf{u}_i\odot\widetilde{\Theta}_i
    \label{formula_lstm_5}
\end{equation}
\begin{equation}
    \bm{h_i}=\mathrm{G}^\mathsf{o}_i\odot\gamma\left(\Theta_i\right)
    \label{formula_lstm_6}
\end{equation}

Formulas~\ref{formula_lstm_1} to \ref{formula_lstm_6} illustrate the computation process after a row vector $\bm{f_i}$ is input into an LSTM cell, where $i \in \mathbb{N}^+$ and $i \leq p$. Here, $\Theta$ denotes the cell state vector, and $\widetilde{\Theta}$ signifies the cell input activation vector. The initial state $\Theta_0$ is initialized as a zero vector with a dimensionality of $h$ (hidden units per LSTM cell). The various gates within the LSTM cell are represented as follows: $\mathrm{G}^\mathsf{f}_i$ is the forget gate, $\mathrm{G}^\mathsf{u}_i$ is the update (or input) gate, and $\mathrm{G}^\mathsf{o}_i$ is the output gate. The weight matrices $\bm{W} \in \mathbb{R}^{h \times m}$ and $\bm{U} \in \mathbb{R}^{h \times h}$, along with the bias vector $\bm{b} \in \mathbb{R}^h$, are parameters learned during the training phase. The activation functions denoted by $\alpha$, $\beta$, and $\gamma$ can be $\mathrm{tanh}$ or $\mathrm{sigmoid}$. The hidden state vector $\bm{h_i}$, which is also the output of the LSTM cell, has a dimensionality of $h$, with the initial hidden state $\bm{h_0}$ being initialized as a zero vector. The operator $\odot$ denotes the Hadamard product (element-wise product)~\cite{horn2012matrix}.

In a bidirectional LSTM, the forward layer generates $\bm{h^f_p}$, while the backward layer produces $\bm{h^b_p}$. The final output of the BiLSTM is obtained by concatenating these two vectors: $\bm{h} = \bm{h^f_p} \oplus \bm{h^b_p}$~\cite{faith1967direct}.

The output of the BiLSTM layer is subsequently passed to the output layer, a multi-label binary classification dense layer, which accepts the vector $\bm{h}$ of size $2h$.

\begin{equation}
    \bm{y} = \mathrm{sigmoid}(\bm{W_c} \bm{h} + \bm{b})
    \label{formula_class}
\end{equation}

In Formula~\ref{formula_class}, binary classification for each category (label) is performed using a $\mathrm{sigmoid}$ activation function. The weight matrix $\bm{W_c} \in \mathbb{R}^{l \times 2h}$, where $2h$ is the size of input vector $\bm{h}$ and $l$ is the number of target labels. The final output is a vector $\bm{y} = [y_1, y_2, \cdots, y_l]$, where each element indicates the probability of the respective intent category. For each element $y_i$, a threshold is applied: $y_i \geq 0.5$ is set to $1$, and $y_i < 0.5$ is set to $0$. This results in a multi-hot vector, where `1' indicates the presence and `0' indicates the absence of an intent. This completes the intent detection process for a smart contract.

\section{Evaluation}
This section outlines the datasets and parameter configurations used in our experiments. We then describe the evaluation metrics and baselines applied. Finally, we present three research questions to comprehensively evaluate our proposed approach.

\subsection{Dataset \& Parameter Configurations}
The dataset is divided into a training set and an evaluation set, each comprising $10,000$ smart contracts. Specifically, the first 10,000 rows are utilized for training, while the rows from $20,000$ to $29,999$ are used for evaluation. The intermediate rows are reserved for potential extended training or evaluation in future experiments.

The training batch size is set to $50$ smart contracts, resulting in a total of $200$ batches to cover the $10,000$ training samples. The models are trained over 100 epochs. The model is compiled using the ``adam'' optimizer and the ``binaryCrossentropy’’ loss function. LSTM models are configured with $64$ hidden units for single layers and $128$ units for two-layered BiLSTM models, a configuration determined optimal based on preliminary experiments. The dimension of the output multi-hot vector $\bm{y}$, as described in Section~\rm{V.C}, is $10$, corresponding to the ten categories of intent we aim to detect.

For intent highlighting using K-means clustering, the initial number of clusters $k$ was set to $19$, based on an occurrence rate threshold of $\rho=0.75$. During training, empty clusters and identical centroids were merged or deleted, refining the number of clusters to $16$. A distance threshold $\lambda=0.21$ was used, with a maximum iteration count set to $80$.

\subsection{Evaluation Metrics}
To evaluate the performance of our model, we utilize the confusion matrix~\cite{powers2020evaluation} and derive several key metrics from it:

\begin{itemize}
    \item \textbf{True Positive (TP)}: The number of correctly predicted intent categories that exist in smart contracts.
    \item \textbf{True Negative (TN)}: The number of correctly predicted non-existing intent categories in smart contracts.
    \item \textbf{False Positive (FP)}: The number of incorrectly predicted intent categories that are falsely identified as existing in smart contracts.
    \item \textbf{False Negative (FN)}: The number of intent categories that exist but are incorrectly predicted as non-existing in smart contracts.
\end{itemize}

Using the values of \textbf{TP}, \textbf{TN}, \textbf{FP}, and \textbf{FN}, we compute the following metrics: \textit{accuracy}, \textit{precision}, \textit{recall}, and \textit{F1-score} (\textit{F1}) as defined in Equations~\ref{formula_accuracy}-\ref{formula_f1}.

\begin{equation}
    \text{accuracy} = \frac{\text{TP} + \text{TN}}{\text{TP} + \text{FP} + \text{FN} + \text{TN}}
    \label{formula_accuracy}
\end{equation}
\begin{equation}
    \text{precision} = \frac{\text{TP}}{\text{TP} + \text{FP}}
    \label{formula_precision}
\end{equation}
\begin{equation}
    \text{recall} = \frac{\text{TP}}{\text{TP} + \text{FN}}
    \label{formula_recall}
\end{equation}
\begin{equation}
    \text{F1} = \frac{2 \times \text{precision} \times \text{recall}}{\text{precision} + \text{recall}}
    \label{formula_f1}
\end{equation}

\textit{Accuracy} is a fundamental metric that indicates the proportion of correct predictions over the total number of predictions.
\textit{Precision} measures the proportion of true positive predictions among all positive predictions made.
\textit{Recall} assesses the proportion of true positive predictions over all actual positive cases in the dataset.
\textit{F1-score} is the harmonic mean of \textit{precision} and \textit{recall}, providing a single metric that balances both concerns.

\subsection{Baselines}
As our study is the first to focus on smart contract intent detection, there are no prior works or experimental results available for direct comparison. 
To address this issue, we perform self-comparisons with several baselines, including classic models such as a basic LSTM model~\cite{hochreiter1997long}, a BiLSTM model~\cite{graves2005framewise2}, and a CNN model~\cite{lecun1995convolutional}. Additionally, we compare our model against advanced generative large language models (LLMs) from OpenAI\footnote{https://openai.com}, specifically GPT-3.5-turbo and GPT-4o-mini, making our evaluation more comprehensive.

We implemented several variants of \textsc{SmartIntentNN} to conduct ablation studies. These variants include an intent highlight model with a scaling factor $\mu=2$, as well as a version without the intent highlight model. Additionally, we replaced BiLSTM with LSTM in certain experimental setups.

\subsection{Research Questions}
We pose three research questions (RQs) to comprehensively evaluate our proposed approach:
\begin{itemize}
    \item \textbf{RQ1:} How effectively does \textsc{SmartIntentNN} detect intents in smart contracts compared to the baselines?
    \item \textbf{RQ2:} How does intent highlight contribute to the performance of \textsc{SmartIntentNN}?
    \item \textbf{RQ3:} How effective is \textsc{SmartIntentNN} in detecting different categories of intents?
\end{itemize}

The above RQs are formulated to analyze from various aspects of \textsc{SmartIntentNN}. RQ1 evaluates its overall performance compared to established baselines. RQ2 investigates the specific contribution of the intent highlight model through ablation test. RQ3 examines its effectiveness across different categories of intents. Next, we will analyze and address these RQs based on the results and visualizations obtained from our evaluation experiments.

\begin{figure}[ht]
    \centering
    \includegraphics[width=\linewidth]{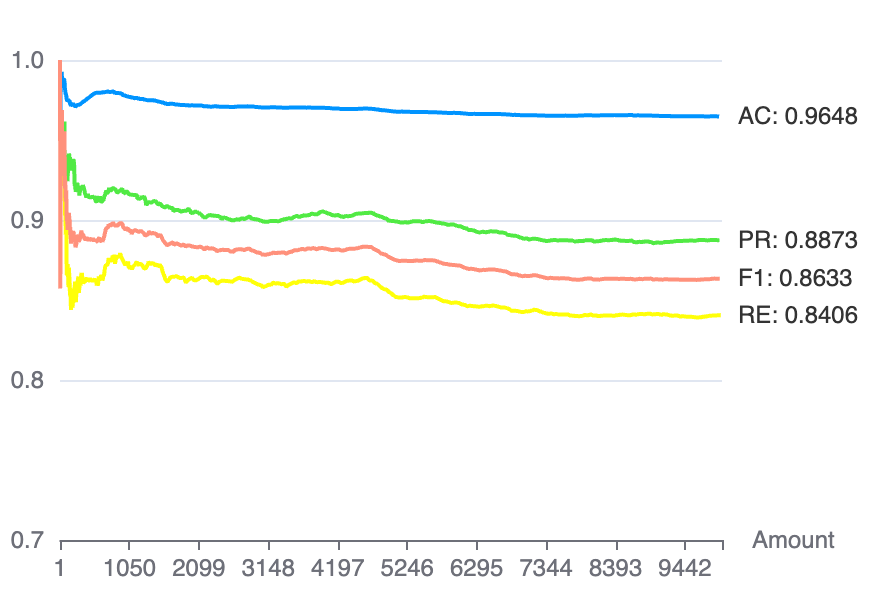}
    \caption{Evaluation metrics trend of \textsc{SmartIntentNN}}
    \label{evaluationtrend}
\end{figure}

\textbf{Firstly, we analyze RQ1.} Fig.~\ref{evaluationtrend} presents a trend chart of the evaluation metrics as the model is evaluated with $10,000$ smart contracts. The metrics \textbf{AC}, \textbf{PR}, and \textbf{RE} represent \textit{accuracy}, \textit{precision}, and \textit{recall}, respectively. It is observed that after approximately $7,344$ evaluations, these metrics stabilize. Therefore, the metrics evaluated on $10,000$ smart contracts are deemed reasonable and reliable. Finally, \textsc{SmartIntentNN} achieves an \textit{F1-score} of $0.8633$, \textit{accuracy} of $0.9647$, \textit{precision} of $0.8873$, and \textit{recall} of $0.8406$, corresponding to the first row of Table \ref{tablebaseline}, which represents the best-performing variant of \textsc{SmartIntentNN}.

\begin{table}[ht]
    \centering
    \caption{Baselines Comparison}
    \begin{tabular}{@{}lcccc@{}}
        \toprule
        \textbf{Model}  & \textbf{Accuracy} & \textbf{Precision} & \textbf{Recall} & \textbf{F1-score} \\
        \midrule
        \multicolumn{5}{c}{\textbf{\textsc{SmartIntentNN} (Ablation Test)}}                            \\
        \midrule
        
        \textbf{USE-$\bm{\mathrm{H_{16}}}$-BiLSTM}  & \small $\bm{0.9647}$  & \small $\bm{0.8873}$  & \small $\bm{0.8406}$  & \small $\bm{0.8633}$     \\
        USE-$\mathrm{H_{2}}$-BiLSTM         & $0.9581$          & $0.8438$           & $0.8386$        & $0.8412$     \\    
        USE-$\mathrm{H_{16}}$-LSTM          & $0.9581$          & $0.8731$           & $0.7999$        & $0.8349$      \\
        USE-BiLSTM                          & $0.9524$          & $0.8337$           & $0.8003$        & $0.8167$      \\
        USE-LSTM                            & $0.9478$          & $0.8319$           & $0.7587$        & $0.7936$      \\
        
        \midrule
        \multicolumn{5}{c}{\textbf{Baseline Models}}                                                      \\
        \midrule
        LSTM            & $0.9172$          & $0.7725$           & $0.5973$        & $0.6737$      \\
        BiLSTM          & $0.9320$          & $0.7871$           & $0.7200$        & $0.7521$       \\
        CNN             & $0.9093$          & $0.6922$           & $0.6596$        & $0.6755$        \\
        GPT-3.5-turbo    & $0.8375$          & $0.4135$           & $0.5447$        & $0.4701$         \\
        GPT-4o-mini    & $0.7821$          & $0.3703$           & $0.9240$        & $0.5288$         \\
        \bottomrule
    \end{tabular}
    \label{tablebaseline}
\end{table}

We further analyze Table \ref{tablebaseline} to compare \textsc{SmartIntentNN} with its baseline models and ablation study variants. The evaluation results demonstrate that \textsc{SmartIntentNN}, incorporating the $\mathrm{H_{16}}$ ($\mu=16$) intent highlight model and BiLSTM, significantly outperforms all ablation study variants and comparative benchmarks. Notable improvements in the \textit{F1-score} include: $28.14\%$ over LSTM, $14.79\%$ over BiLSTM, $27.80\%$ over CNN, $83.64\%$ over GPT-3.5-turbo, and $63.26\%$ over GPT-4o-mini. Furthermore, ablation tests reveal that \textsc{SmartIntentNN} with a BiLSTM layer outperforms the single LSTM variant by $3.40\%$ in \textit{F1-score}, underscoring the advantage of BiLSTM in capturing contextual information from both past and future data, thus enabling a better understanding of the semantic structure for more accurate intent classification.

Specifically, \textsc{SmartIntentNN} outperforms LLM baselines because LLMs, such as OpenAI's GPTs, are designed for general-purpose tasks. Although they possess strong generalization capabilities, they lack the fine-tuning required for the specific domain of smart contract intent detection. In contrast, \textsc{SmartIntentNN} is explicitly trained for this task, enabling it to capture and understand the contextual semantics of the code and the specific intents of the developers. Therefore, surpassing LLMs is an expected outcome.

\begin{tcolorbox}
\textbf{Answer to RQ1} $\blacktriangleright$
By utilizing a USE pre-trained model, a K-means-based intent highlight model, and a BiLSTM-based multi-label classification DNN, \textsc{SmartIntentNN} can effectively detect and classify ten types of unsafe development intents in smart contracts, outperforming classic deep learning models and cutting-edge LLMs.
\end{tcolorbox}

\begin{figure}[ht]
    \centering
    \subfigure[All K-means clusters]{\includegraphics[width=0.48\linewidth]{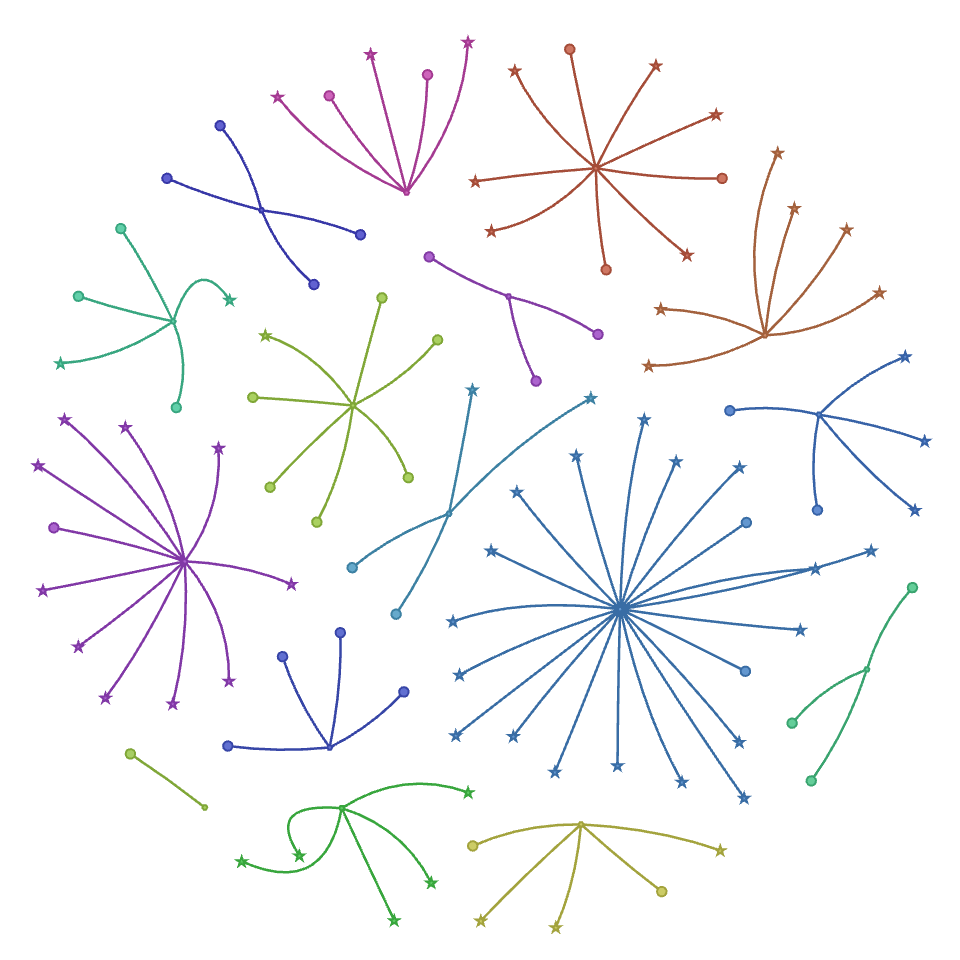}}
    \hspace{0.01\linewidth}
    \subfigure[Example of one cluster]{\includegraphics[width=0.48\linewidth]{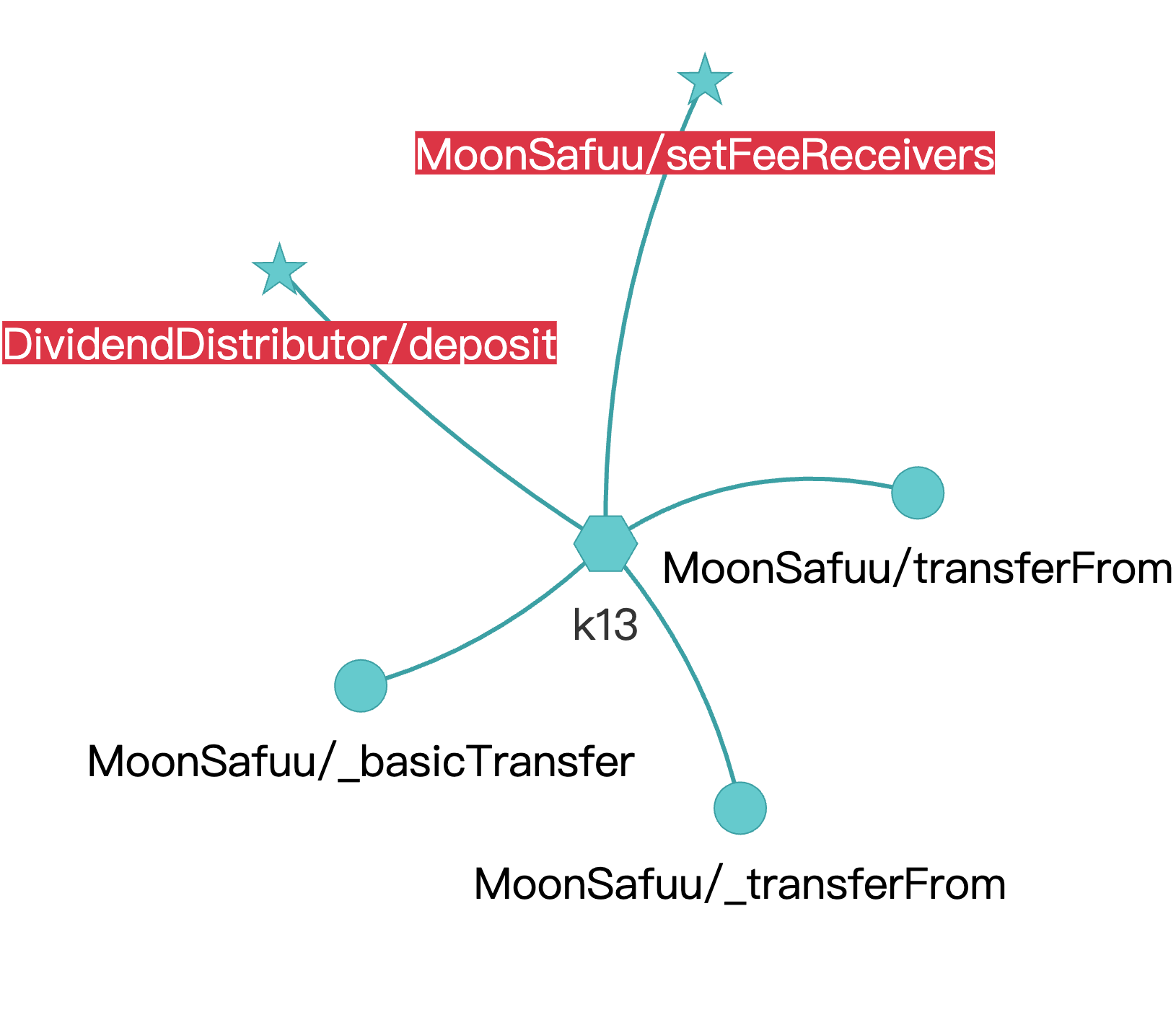}}
    \vfill
    \subfigure[Detailed visualization of highlighted smart contract functions]{\includegraphics[width=\linewidth]{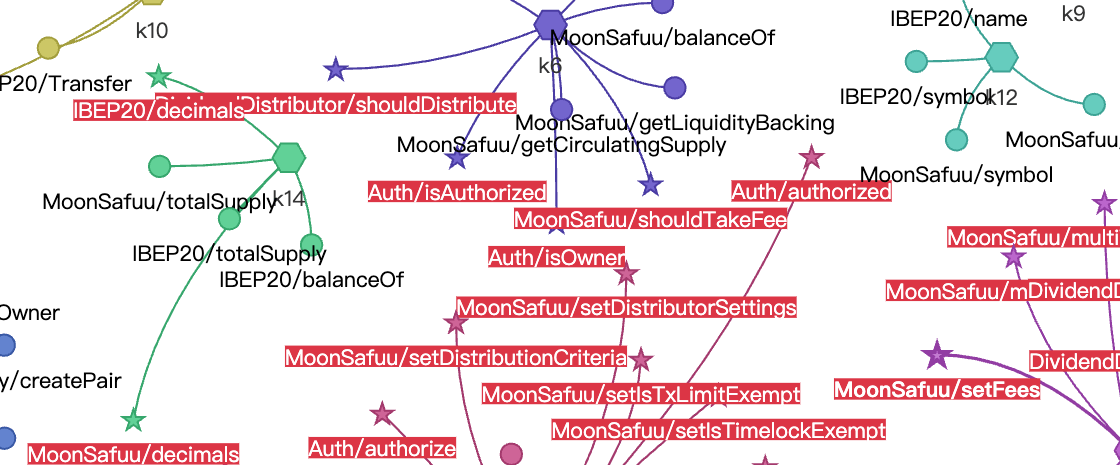}}
    \caption{Visualization of the Intent Highlight Model}
    \label{visualintent}
\end{figure}

\textbf{Secondly, we analyze RQ2.}
The intent highlight model identifies and emphasizes distinctive \textit{functions} with strong intents through K-means clustering. As shown in Fig.~\ref{visualintent}, Subfigure (a) displays all clusters, where each of the $16$ clusters represents a common \textit{function}'s spatial centroid. Subfigure (b) focuses on one cluster, where hexagons mark centroids, circles represent \textit{functions} with weaker intents, and stars highlight \textit{functions} with stronger intents. Edges between nodes indicate within-cluster distances, illustrating intent strength. Subfigure (c) highlights various \textit{functions}, with red labels identifying those with strong intents, such as \textit{setFees} and \textit{setIsTxLimitExempt}. While these highlighted \textit{functions} may not be inherently malicious, they are distinctive enough to be emphasized. The vector features of these \textit{functions} are then scaled by a factor of $\mu$ to enhance learning in subsequent DNN layers.

From Table \ref{tablebaseline}, it is clear that \textsc{SmartIntentNN} variants incorporating the intent highlight model offer substantial performance improvements. Specifically, with LSTM, the variant using intent highlight model $\mathrm{H_{16}}$ ($\mu=16$) shows a $5.20\%$ gain in \textit{F1-score} over its non-highlight counterpart. Similarly, for BiLSTM variants, introducing $\mathrm{H_{2}}$ ($\mu=2$) results in a $3.00\%$ increase, and this improvement escalates to $5.71\%$ with $\mu=16$. These findings confirm the substantial impact of the intent highlight model in improving performance.

\begin{tcolorbox}
\textbf{Answer to RQ2} $\blacktriangleright$
The intent highlight model markedly enhances \textsc{SmartIntentNN}'s performance in intent detection by amplifying features of \textit{functions} with prominent intents, making the learning process more robust and accurate.
\end{tcolorbox}

\begin{figure}[ht]
    \centering\includegraphics[width=\linewidth]{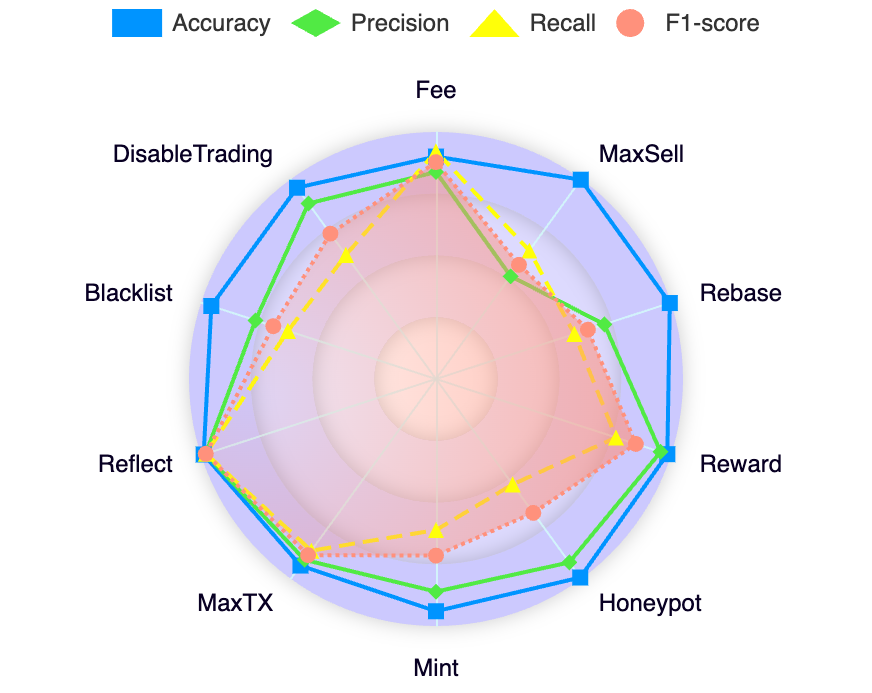}
    \caption{
        Evaluation metrics of \textsc{SmartIntentNN} for different intents.
    }
    \label{evaluationradar}
\end{figure}

\textbf{Finally, we analyze RQ3.} As shown in Fig.~\ref{evaluationradar}, the radar chart illustrates the performance metrics for ten distinct intent categories. Specifically, red circles represent the \textit{F1-score}, green diamonds indicate the \textit{precision}, yellow triangles denote the \textit{recall}, and blue rectangles signify the \textit{accuracy}. The ten axes on the radar chart correspond to the ten different intent categories. From the chart, we find that the performance of \textsc{SmartIntentNN} in detecting different categories of intents varies.

From the radar chart, it is evident that three intents, \textbf{Reflect}, \textbf{MaxTX}, and \textbf{Fee}, achieve the highest \textit{F1-scores}, with values of $0.98$, $0.88$, and $0.87$ respectively, demonstrating outstanding detection performance. Conversely, the \textit{F1-score} for \textbf{MaxSell} is the lowest at $0.57$, indicating relatively poor performance in detecting this intent. This discrepancy can be attributed to the imbalance in our dataset. The proportion of \textbf{Reflect}, \textbf{MaxTX}, and \textbf{Fee} intents is significantly higher, allowing the model to be more effectively trained on these categories. In contrast, \textbf{MaxSell} samples constitute only $0.05\%$ of the dataset, leading to insufficient training and consequently lower detection performance.

\begin{tcolorbox}
\textbf{Answer to RQ3} $\blacktriangleright$
\textsc{SmartIntentNN} exhibits varying effectiveness in detecting different intent categories. Although detection is weaker for certain intents, the overall accuracy for each category remains high, exceeding $90\%$, underscoring the model's efficacy in distinguishing intent presence or absence. As data collection becomes more balanced, the model's performance across different intents is expected to equalize.
\end{tcolorbox}

\section{Threats to Validity}
\subsection{Internal Validity}
A major threat to the internal validity of our study is the dataset imbalance. For example, the \textbf{MaxSell} intent constitutes only $0.05\%$ of our dataset, considerably less than other intent labels. This scarcity could limit our model's ability to accurately identify this intent. To address this, we employ techniques such as oversampling, undersampling, and cross-validation to mitigate the imbalance. However, the natural rarity of these intents reflects their real-world prevalence, suggesting proportionately lower associated risks. Additionally, \textsc{SmartIntentNN} is designed to be continuously trainable; as we gather more data on these rarer intents, we will retrain and fine-tune the model to improve detection performance. Despite these measures, the detection of less frequent intents like \textbf{MaxSell} remains less effective, but this mirrors their limited impact in practical scenarios.

\subsection{External Validity}
The principal challenge to the external validity of our model stems from the diversity of programming languages used for smart contract development. Primarily trained on Solidity, the most popular smart contract programming language~\cite{top10lanaguge2019blockchain}, our model effectively covers a vast majority of real-world smart contracts. While our current model does not yet extend to other languages (e.g., Vyper~\cite{buterin2018vyper} and JavaScript), the adaptability of \textsc{SmartIntentNN} allows it to be retrained with additional data to support these languages. For bytecode-only contracts, deploying decompilers to convert the bytecode to source code allows the model to predict intents based on the contextual semantics embedded in the contracts. By virtue of its continuous retraining capability, \textsc{SmartIntentNN} maintains robustness and effectiveness, ensuring it can adapt to the evolving landscape of smart contract languages and effectively address external validity.

\section{Related Work}
In this section, we review related work on external and internal risks and explain how our approach differentiates itself from previous studies.

\subsection{External Risks}
Traditional formal methods have significantly contributed to smart contract vulnerability detection.
%Traditional formal methods have made significant contributions to the detection of vulnerabilities in smart contracts.
Tools like Oyente~\cite{luu2016making} and Mythril~\cite{mueller2017framework} use symbolic execution and control flow verification. Vandal~\cite{brent2018vandal} employs logical specifications for EVM bytecode analysis, while ZEUS~\cite{kalra2018zeus} leverages abstract interpretation and symbolic model checking. SmartCheck~\cite{tikhomirov2018smartcheck} and Securify~\cite{tsankov2018securify} utilize static analysis techniques. \textsc{TeEther}~\cite{krupp2018teether} automates vulnerability identification and exploit generation directly from binary bytecode. Additional methods include sFuzz~\cite{nguyen2020sfuzz}, which employs branch distance-driven fuzzing, Osiris~\cite{torres2018osiris}, combining symbolic execution and taint analysis for integer error detection, Slither~\cite{feist2019slither}, which uses static analysis to convert smart contract code into an intermediate representation for vulnerability detection, and SMARTIAN~\cite{choi2021smartian}, integrating static and dynamic analyses to effectively discover bugs in smart contracts. With the advancement of deep learning, detection efficiency and accuracy have improved. SaferSC~\cite{tann2018towards} employs LSTM-based sequential learning, and ContractWard~\cite{wang2020contractward} uses machine learning for feature extraction. DR-GCN and TMP~\cite{zhuang2020smart} integrate graph neural networks, while CGE~\cite{liu2021combining} and AME~\cite{liu2021smart} combine deep learning with expert patterns. ESCORT~\cite{sendner2023smarter} uses multi-label classifiers and transfer learning, and DMT~\cite{qian2023cross} enhances bytecode vulnerability detection. SCVHunter~\cite{luo2024scvhunter} utilizes heterogeneous graph attention networks, and Clear~\cite{chen2024improving} leverages contrastive learning to significantly improve detection performance.

All the above methods focus on detecting vulnerabilities as external risks caused by development bugs. However, our work targets internal risks by detecting intentionally unsafe code, filling a gap in smart contract security study.

\subsection{Internal Risks}
Internal risks, with fewer studies, often result from developers' malicious intent.
Phishing scams have been investigated in several studies, including those by SIEGE~\cite{li2023siege} and DElightGBM~\cite{chen2020phishing}. \textsc{HoneyBadger}~\cite{torres2019art} exposes honeypots through symbolic execution, while \textsc{FairCon}~\cite{liu2020towards} assesses contractual fairness in smart contracts. The “Trade or Trick” approach~\cite{xia2021trade} employs machine learning techniques to detect scam tokens on the Uniswap platform.
SCSGuard~\cite{hu2022scsguard} targets scam detection in bytecode, specifically addressing Ponzi schemes and honeypots. More recent advances include TTG-SCSD~\cite{fan2022smart}, which employs topological data analysis for detecting smart contract scams, and CryptoScamTracker~\cite{li2023double}, which addresses cryptocurrency giveaway scams using certificate transparency logs. Pied-Piper~\cite{ma2023pied} integrates datalog analysis with directed fuzzing to uncover backdoors in Ethereum ERC token contracts. Additionally, Tokeer~\cite{zhou2024stop} investigates rug pull risks by leveraging configurable transfer models.

Although these studies address various aspects of internal risks, such as phishing scams, rug pulls, and Ponzi schemes, none systematically define or detect a wide range of unsafe developer intents.
In contrast, our study identifies ten specific negative intents and employs \textsc{SmartIntentNN} for comprehensive multi-label classification, offering a more robust solution to internal risk analysis.

\section{Conclusion}
We have proposed \textsc{SmartIntentNN}, the first model for detecting development intent in smart contracts. The model comprises a pre-trained USE model, an intent highlight model based on K-means clustering, and a DNN integrated with a BiLSTM layer. We trained and evaluated our model on a dataset of over $40,000$ smart contracts, which were cleaned and labeled into ten distinct categories of intents. Experimental results demonstrate that our model surpasses all baselines, achieving an \textit{accuracy} of $0.9647$, a \textit{precision} of $0.8873$, a \textit{recall} of $0.8406$, and an \textit{F1-score} of $0.8633$. Our work fills a critical gap in smart contract security study by addressing developer intent detection.

\section*{Acknowledgment}
This work was supported by the Macao Science and
Technology Development Fund under Grant 0161/2023/RIA. 

%\clearpage
\bibliographystyle{IEEEtran}
\bibliography{IEEEabrv,refs}

\end{document}